# Creating bio-inspired hierarchical 3D-2D photonic stacks via planar lithography on self-assembled inverse opals

**Ian B. Burgess[1,4], Joanna Aizenberg[1,2,3], and Marko Lončar[2]**

[1]Wyss Institute for Biologically Inspired Engineering, Harvard University, Cambridge MA, USA,
[2]School of Engineering and Applied Sciences, Harvard University, Cambridge MA, USA
[3]Department of Chemistry and Chemical Biology, Harvard University, Cambridge MA, USA
[4]To whom correspondence should be addressed

Email: ibburges@fas.harvard.edu

**Abstract.** Structural hierarchy and complex 3D architecture are characteristics of biological photonic designs that are challenging to reproduce in synthetic materials. Top-down lithography allows for designer patterning of arbitrary shapes, but is largely restricted to planar 2D structures. Self-assembly techniques facilitate easy fabrication of 3D photonic crystals, but controllable defect-integration is difficult. In this paper we combine the advantages of top-down and bottom-up fabrication, developing two techniques to deposit 2D-lithographically-patterned planar layers on top of or in between inverse-opal 3D photonic crystals and creating hierarchical structures that resemble the architecture of the bright green wing scales of the butterfly, *Parides sesostris*. These fabrication procedures, combining advantages of both top-down and bottom-up fabrication, may prove useful in the development of omnidirectional coloration elements and 3D-2D photonic crystal devices.

**1. Introduction**

Top-down nanofabrication based on planar lithography has been the workhorse of modern nanotechnology, enabling the current microelectronics industry as well as the development of a wide range of platforms (photonics, microfluidics, etc.). However, while planar lithography allows the designer sculpting of a wide range of materials in 2D, the extension of such capabilities to truly 3D architectures has proved challenging [1]. In contrast, 3D nanostructures with complex hierarchical architecture are ubiquitous in biological systems, and are often tailor-made for a specific function (e.g. omnidirectional coloration, superhydrophobicity) [2]. The ability to tailor the shape of materials in 3D would open the door for many exciting technologies, including several that have already been proposed. A prominent example in photonics, the prospect of 3D-photonic crystals (dielectric superlattices with wavelength-scale periodicity) with designer defects has the promise to enable devices that channel the flow of light without any losses [1,3-7]. This could, for example, lead to the development of defect-tolerant ultrahigh-Q optical resonators.

Presently, there are very few techniques that allow truly 3D nanofabrication, and those that do exist are currently limited in their applicability. Focused ion-beam milling can be used to mold a wide variety of materials in 3D, however, this technique is both not scalable (very time consuming, cannot be done in parallel), and generally results in material damage due to ion implantation [8]. Direct-laser writing [1] and two-photon lithography [9] can be used to imprint designer 3D patterns in appropriate resist materials (these patterns can be transferred to other materials by infiltration and inversion), however their resolution is limited by the wavelength of light, which also defines the relevant feature-sizes for photonic-crystal devices.

Bottom-up techniques based on self-assembly have emerged as a powerful platform for templating structures that are patterned in 3D with feature sizes ranging from a few nanometers to several microns [10-19]. Unlike their top-down counterparts, they allow scalable 3D patterning of materials. However the variety of patterns that can be made through bottom-up techniques are very limited in comparison to top-down lithography (generally restricted to periodic or polycrystalline patterns with random defects). For example, while self-assembly has enabled the realization of 3D photonic crystals with complete photonic band-gaps [20], the incorporation of designer defects has thus far been very limited [21], preventing the experimental realization of many of the device properties that initially generated excitement about 3D photonic crystals [3].

In this paper we develop techniques to build stacks of self-assembled porous 3D-photonic crystals and lithographically patterned planar defect layers with comparable feature-sizes to the pores. This type of structural hierarchy is found in the bright green wing scales of the butterfly, *Parides sesostris*. In this organism, the 2D layer serves as a diffuser, enhancing the omnidirectionality of the photonic-crystal's iridescent color [22]. Recent proposals [4-6] have also shown that many of the attractive properties of 3D photonic crystal circuits (such as diffractionless control of light flow) can also be accomplished in hybrid 3D-2D-3D structures, consisting of 2D photonic crystals with designer defects sandwiched between two uniform 3D photonic crystals. We developed two techniques to incorporate lithographically patterned planar layers on top of or inside of self-assembled 3D porous inverse-opal films.

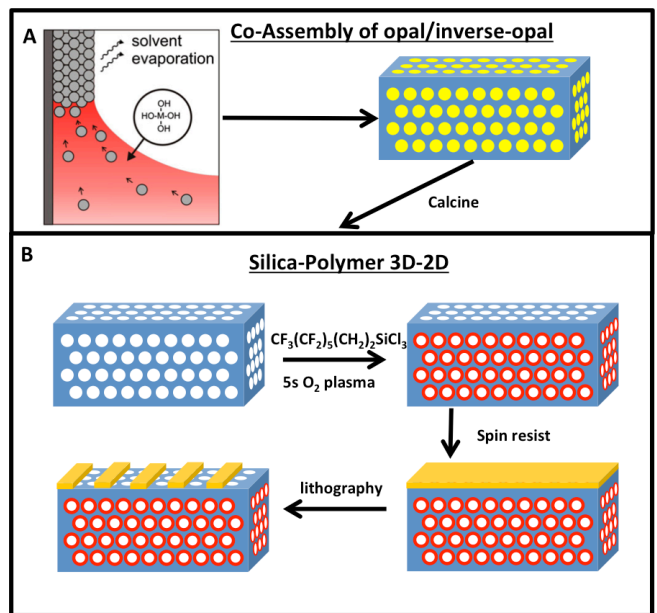

**Figure 1**. Exploitation of selective wetting to facilitate the fabrication of 2D-planar structures on top of porous inverse-opal films (IOFs). After standard IOF fabrication, the porous silica network is functionalized with (1H,1H,2H,2H-tridecafuorooctyl) trichlorosilane (13FS) to facilitate prevention of solvent and resist infiltration. 5s of oxygen plasma is applied to make the top-surface sufficiently adhesive for the resist, while preserving the 13FS functionality further down the structure. A resist layer can then be spun and patterned using standard lithographic techniques.

**2. Selective wetting as a means to separate planar and porous films**

Previously, we developed a colloidal-co-assembly process to produce inverse-opal films (IOFs) with flat top-surfaces and exceptional long-range order [23]. We then developed techniques to pattern the surface chemistry in the pores these structures and showed that this gave us exceptional control as to where specific liquids would be able to wet the structures [24-26]. We exploit this selective wetting to direct the deposition of resist on the top of the IOFs without penetrating the porous structure. The fabrication protocol is shown in Fig. 1. IOFs are functionalized first by exposure to vapors of (1H,1H,2H,2H-tridecafuorooctyl) trichlorosilane (13FS). These groups provide the structure with the ability to resist infiltration of nearly any liquid, including the precursors and solvents for most resists. We then exposed to the surface to oxygen plasma for a very short time (5s, 100W, 5-10sccm oxygen flow), much shorter than would be used even to form a significant vertical gradient of wettability in the structure [24-26]. The purpose of this step is to activate the top surface just enough for sufficient adhesion of the resist to occur, while still not enough to let the resist penetrate beyond the first row of inter-pore necks [25-26]. Finally the resist can be spun and patterned. Fig. 2A shows a cross-section of an IOF with a film of SU-8 photoresist spun on top of it (SU-8 2002, MicroChem, spun at 3000 rpm). While delamination was not observed, even during cleaving, indicating some adhesion between the film and the substrate, penetration of the SU-8 is not observed beyond the top layer of half-pores. In particular, no resist is observed beyond the top row of necks. In contrast, when IOFs are not functionalized in the above manner, the SU-8 resist easily penetrates the pores and plugs the structure, as seen in Fig. 2B. Fig. 2C,D show a patterned film of ZEP520 resist (ZEP520-A, Zeon corp., patterned by direct electron-beam lithography) suspended on top of an IOF. The resist was deposited from an anisole solution. Again, adhesion of the resist to the film was observed, but no infiltration of the pores beyond the top row of necks.

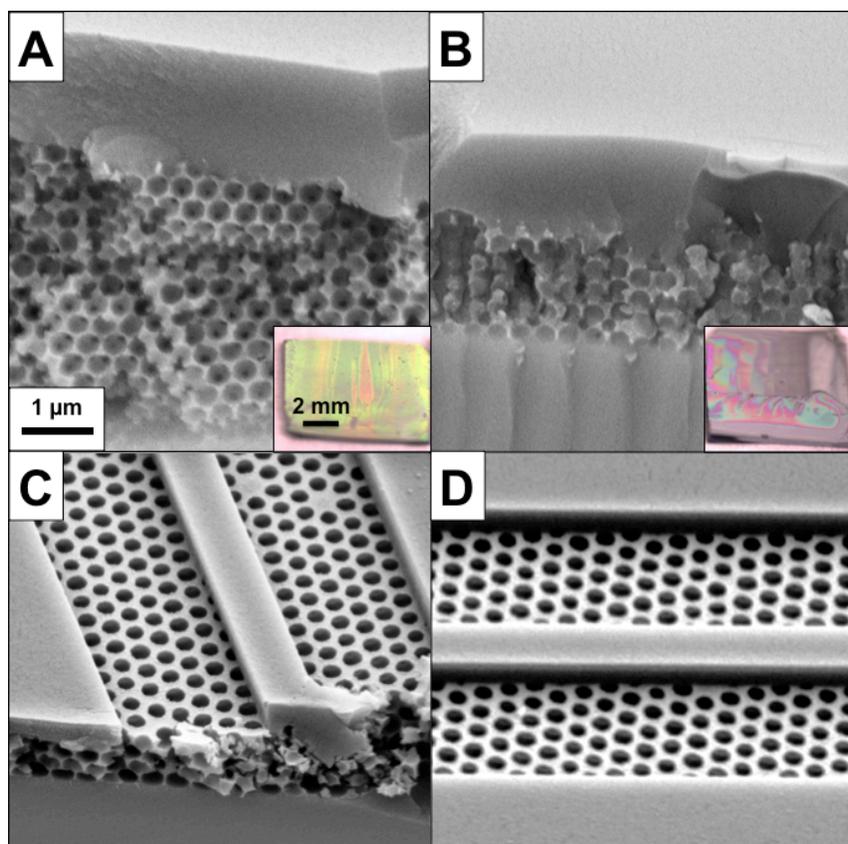

**Figure 2**. (A) Cross-section showing a planar SU-8 film deposited on top of a functionalized IOF. The SU-8 does not penetrate the pores beyond the top layer of half-spheres. (B) Cross-section of an unfunctionalized IOF after spin-coating SU-8, showing penetration of the resist into the porous network. The insets in A and B show optical images of the films after SU-8 deposition. The iridescent green color form the IOF is quenched on the right due to index

matching with the resist that has filled the pores. (C,D) SEM images of a film of ZEP520 resist, patterned by electron-beam lithography, suspended on top of an IOF.

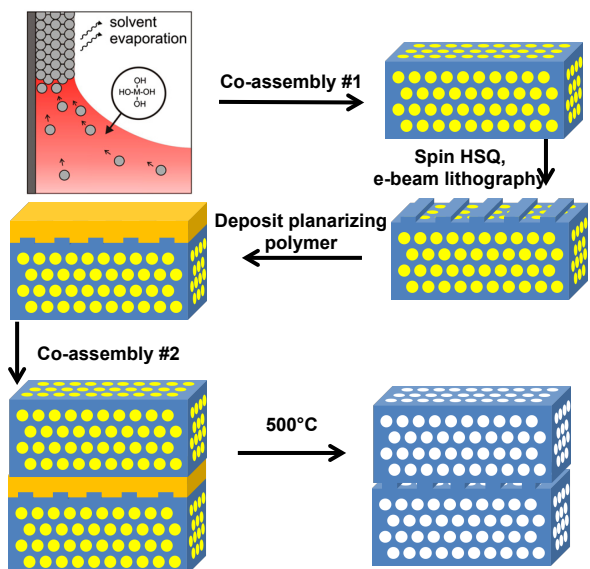

**Figure 3**. Schematic showing the plugged-IOF route to stacking of IOF and 2D-patterned planar layers. Co-assembly of the polymer colloidal crystal template and the silica interstitial matrix is performed [23], but the polymer spheres are not immediately removed, resulting in a solid, non-porous film (plugged IOF). A spin-on-glass electron-beam resist (hydrogen silsesquioxane, HSQ) is deposited on top of the plugged IOF and patterned by electron-beam lithography. A thick polymer film is then spun on top of the 2-D patterned layer to planarize. Following the activation of the planarizing film surface, a second co-assembly step is done to deposit a second plugged IOF. Finally all polymer components are removed by a slow calcination procedure (ramp up to 500°C over 5hr), leaving behind a stack consisting of a 2D-patterned planar layer wedged between two IOFs.

**3. 3D-2D-3D silica stacks from plugged inverse-opal films and planarized 2D layers**

While possessing the highly selective wetting properties that enabled the previous fabrication procedure, IOFs fabricated by co-assembly also benefit from the advantage of having very flat top surfaces after deposition [23]. Before removal of the polymer template (a state referred to as a plugged IOF hereafter), plugged IOFs are completely flat and solid films of material [23]. Owing to the flatness and lack of porosity in plugged IOFs, resist layers can be easily deposited and patterned before removal of the polymer spheres. However, this order of fabrication steps requires the use of a resist that can withstand the conditions used to remove the polymer spheres (this rules out many organic polymer-based resists). We used spin-on-glass (hydrogen silsesquioxane, HSQ) as a resist for this demonstration since upon curing it becomes chemically similar to the sol-gel silica material that makes up the IOF. The resist was patterned by electron-beam lithography. When the structure was calcined after this e-beam patterning, a 3D-2D hybrid structure similar to the type formed in the previous section is left behind (Fig. 4A).

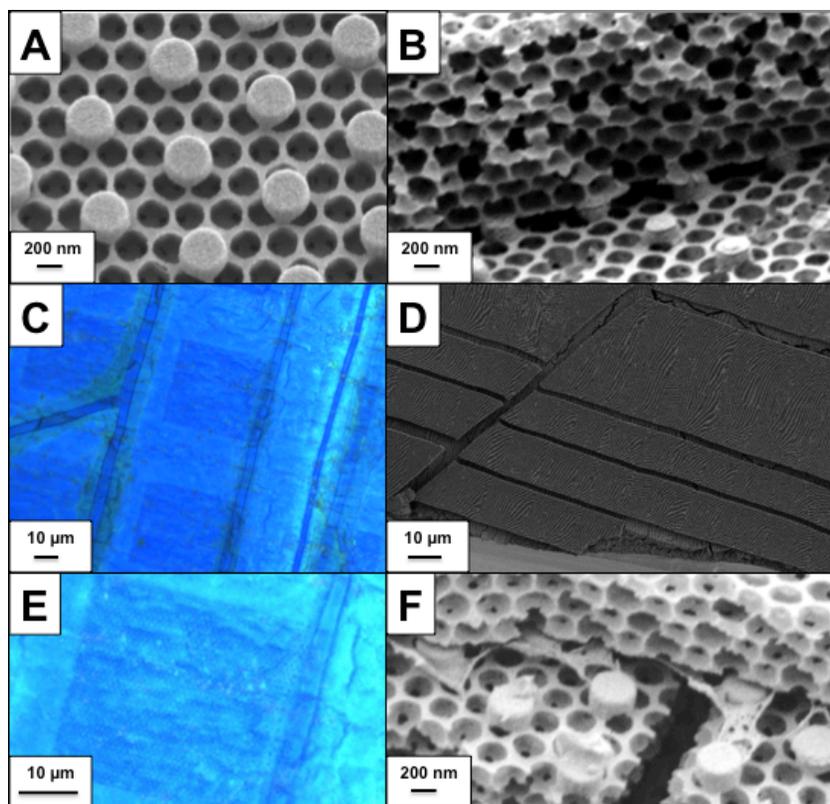

**Figure 4**. (A,B) SEM images 3D-2D and 3D-2D-3D stacks prepared from plugged IOFs. (A) A 3D-2D structure is formed by removing the polymer template right after the electron-beam lithography step. An array of 500nm wide posts are shown suspended on top of an IOF with a ~300nm pore diameter. (B) A cross-section of a 3D-2D-3D structure showing adhesion between the layers. The same array of 500nm wide posts is used in this structure. (C,D) Large-area optical (C) and scanning-electron (D) microsgaphs of 3D-2D-3D stacks. (E) Higher-magnification optical micrograph, showing the bright coloration with the 2D array of posts visible underneath the first 3D layer. (F) SEM cross-section showing that while SU-8 served as a planarizing layer that remained hydrophilic after activation for a sufficiently long time to allow deposition of IOFs, it had the tendency to leave behind residue after calcination.

However, if the IOF remains plugged, a second IOF can be grown on top of the patterned HSQ layer, sandwiching this lithographically 2D-patterned layer between two 3D photonic crystals, producing the type of 3D-2D-3D architecture proposed in Refs. [3-4]. Successful deposition of the second IOF requires first planarizing the patterned 2D HSQ layer so that the nanostructures do not interfere with the growth of the second IOF. Spin-coating a thick film on top of the HSQ pattern can accomplish this. We deposited a planarizing film with ~5μm thickness for a ~250nm thick resist layer. In order for the second IOF to grow successfully, the surface of this planarizing layer must be able to be rendered very hydrophilic (zero-degree contact angle) for a long time (typical IOF deposition takes 1.5 days).

We found that many common polymers were not suitable substrates for a successful deposition of a second IOF, even after surface activation by oxygen plasma (we tried poly methyl-methacrylate, polystyrene, and polypyrrole with no success), we found that growth of good-quality IOFs was possible on films of cured SU-8 (flood exposed with UV-light to cure). After electron-beam patterning of the resist layer and development, a planarizing SU-8 film with ~5μm thickness (SU-8 2005, MicroChem, spun at 3000rpm) was deposited and flood exposed. The film was then rendered hydrophilic via a 2min exposure to oxygen plasma (100W, 5-10 sccm oxygen flow). The second IOF was then deposited on top of this film. After deposition of the second plugged IOF, the entire structure was calcined to remove the polymer spheres and the planarizing polymer in a single step. To facilitate gentle removal of the planarizing polymer, allowing for the top layer IOF to fall on top of the lower layers without large amounts of cracking, the temperature was ramped up to 500°C over 5 hrs (held for 2 hrs). Before deposition of the second IOF, it is also possible to thin down the planarizing layer using reactive-ion etching. However this process, if not strictly necessary, would add significant complexity to the otherwise simple fabrication procedure.

Fig. 4B shows a cleaved section of a 3D-2D-3D stack formed using an SU-8 planarization layer. As evident from the cross section, the top layer IOF successfully adhered to the surface of the 2D layer after calcining. Figs. 4C and 4D show low-magnification images of the stacks. While we observed a higher density of cracks on the top IOF than in the bottom one, large crack-free stacked regions could still be made. The single preferred crystal orientation of the IOFs [23] was also maintained. This is clearly illustrated in the oriented crack structure visible in Fig. 4D. Shown in Fig. 4C,E, a 30 μm x 30 μm array of posts in the 2D layer can easily be contained in a crack-free region. Although SU-8 was found to be a polymer planarizing layer that could remain activated for sufficiently long times to allow for successful deposition of the top layer IOF, we did find that the calcination process sometimes left behind a noticeable amount of residue in the 2D layer. Fig. 4D shows examples of this residue in a cleaved stack. Due to the large quantity of resist in the planarizing layer and the highly aromatic structure of SU-8, this residue is likely carbonaceous remnants of calcination.

## 4. Conclusion

In this paper, we have demonstrated two techniques to integrate 2D-lithographically-patterned planar layers on top of or in between 3D-periodic inverse-opal films. Both of these techniques are enabled by the uniform, crack-free and flat-surfaced nature of these films when grown by colloidal co-assembly [23]. Using simple chemical surface-modifications and the unique wetting properties of the IOFs (these properties are characterized in ref [25]), we have shown that we are able to spin-coat and pattern resist layers on top of IOFs that show good adhesion to the top of the porous surface, but no penetration inside of the pores. Using plugged IOFs as flat solid films, we were able to stack alternating layers of 3D-periodic IOFs and 2D-periodic planar spin-on-glass layers before removing the polymer fillings in a single step. These fabrication procedures, combining advantages of both top-down and bottom-up fabrication, may lead to the development of omnidirectional color elements inspired by the wing-structure of the butterfly, *Parides sesostris* [22]. This type of structure may also serve as a useful template for the deposition of high-refractive-index inorganics. This could lead to the development of 3D-2D photonic crystals that exhibit diffractionless confinement of light in complex networks of designer defects [3-5].

**Acknowledgements**

We thank L. Mishchenko, B.D. Hatton, N. Koay, and B.A. Nerger for helpful discussions. This work was supported by the AFOSR Award # FA9550-09-1-0669-DOD35CAP. IBB acknowledges support from the Natural Sciences and Engineering Research Council of Canada through the PGS-D program.